%% file: ontology.tex
\documentclass[sigconf]{acmart}

\usepackage{booktabs} 

\setcopyright{acmcopyright}

\acmConference[Ninth International Conference on Knowledge Capture 2017]{K-CAP 2017}{December 2017}{Austin, Texas} 

\usepackage{graphicx}
\usepackage{mathtools}

\usepackage{amsmath,amssymb,amsfonts}

\usepackage{url}
\usepackage{verbatim}
\usepackage{multirow}
\usepackage{caption}
\usepackage{subcaption}

\usepackage[linesnumbered,ruled,vlined]{algorithm2e}
\SetKwInput{KwInput}{Input}
\SetKwInput{KwOutput}{Output}
\SetKwProg{myproc}{Procedure}{}{}

\usepackage{algpseudocode}
\usepackage{mathtools}
\usepackage{amsmath}

\usepackage{ifpdf}
\usepackage{lmodern}
\usepackage{lipsum}
\usepackage[T1]{fontenc}
\usepackage{textcomp}
\usepackage{enumitem}
\usepackage{mdwlist}
\usepackage{pbox}
\usepackage{comment}
\usepackage{ifthen}
\usepackage{color}
\usepackage{colortbl,xcolor}
\usepackage{upgreek}
\usepackage{upgreek}
\usepackage{todonotes}
\usepackage{breakurl}

\usepackage{tikz}
\usetikzlibrary{arrows}

\usepackage[para]{footmisc}

\definecolor{Gray}{gray}{0.9}
\definecolor{LightCyan}{rgb}{0.88,1,1}

\specialcomment{notes}{\begingroup \color{blue}} { \endgroup}

\newcolumntype{!}{>{\global\let\currentrowstyle\relax}}
\newcolumntype{^}{>{\currentrowstyle}}

\newcommand{\si}{\begin{enumerate}}

\newcommand{\ei}{\end{enumerate}}

\makeatletter
\let\oldfootnote\footnote
\def\footnote{\@ifstar\footnote@star\footnote@nostar}
\def\footnote@star#1{{\let\thefootnote\relax\footnotetext{#1}}}
\def\footnote@nostar{\oldfootnote}
\makeatother

\DeclareGraphicsExtensions{.eps,.png}
\graphicspath{{figure/}{figure/eps/}{figure/png/}}

\newcommand{\superscript}[1]{\ensuremath{^{\textrm{#1}}}}

  \renewenvironment{thebibliography}[1]{%
    \begin{oldthebibliography}{#1}%
      \setlength{\parskip}{0ex}%
      \setlength{\itemsep}{0ex}%
  }%
  {%
    \end{oldthebibliography}%
  }
  
\begin{document}

\title[Klout Topic Ontology]{Klout Topics for Modeling Interests and Expertise of Users Across Social Networks}

\def\kloutinc{\superscript{*}}
\def\rutgers{\superscript{\dag}}

\author{Sarah Ellinger, Prantik Bhattacharyya, Preeti Bhargava, Nemanja Spasojevic}
\affiliation{%
  \institution{Lithium Technologies | Klout}
  \city{San Francisco} 
  \state{CA} 
}
\email{{sarah.ellinger, prantik.bhattacharyya, preeti.bhargava, nemanja.spasojevic}@lithium.com}

%
%

\renewcommand{\shortauthors}{S. Ellinger et. al.}

\begin{abstract}
\input{texfiles/ontology_abstract}
\end{abstract}



\maketitle

\section{Introduction}
\label{section:introduction}
\input{texfiles/ontology_introduction}


\section{Motivation}
\label{section:problem_statement}
\input{texfiles/ontology_problem_statement}

\section{Best Practices}
\label{section:best_practices}
\input{texfiles/ontology_best_practices}

\section{Alternative Topic Sets}
\label{section:ontology_alternatives}
\input{texfiles/ontology_alternatives}

\section{Klout Topics Characteristics}
\label{section:ontology_characteristics}
\input{texfiles/ontology_characteristics}

\section{Klout Topics Coverage Versus Alternatives}
\label{section:ontology_results}
\input{texfiles/ontology_results}

\section{Case Studies}
\label{section:ontology_sample_case_study}
\input{texfiles/ontology_sample_case_study}

\section{Conclusion and Future Work}
\label{section:conclusion}
\input{texfiles/ontology_conclusion}

\vspace{-0.03in}



\bibliographystyle{ACM-Reference-Format}
\bibliography{ontology_bibliography.bib}

\end{document}

%% file: texfiles/ontology_abstract.tex
This paper presents Klout Topics, a lightweight ontology to describe social media users' topics of interest and expertise. Klout Topics is designed to: be human-readable and consumer-friendly; cover multiple domains of knowledge in depth; and promote data extensibility via knowledge base entities. We discuss why this ontology is well-suited for text labeling and interest modeling applications, and how it compares to available alternatives. We show its coverage against common social media interest sets, and examples of how it is used to model the interests of over 780M social media users on Klout.com. Finally, we open the ontology for external use.

%% file: texfiles/ontology_introduction.tex

The richness of social media as a source of information about its users' interests and areas of expertise has led to a great deal of research on natural language processing (NLP) algorithms for extracting that data. Taking free-form text and modeling it as structured topics opens up a variety of applications that are not possible with unstructured bags of keywords. However, an often-neglected aspect of the modeling task is choosing \textit{which} set of candidate topics to use. Through the practice of modeling social media interest and expertise topics on Klout.com, we have identified several criteria for such a topic set. These include: 
\begin{itemize}[noitemsep,nolistsep]
\item covering many domains of knowledge in depth and in multiple languages; 
\item being human-readable, so that modeled topics can be exposed by an application;
\item being small enough for human curation to be feasible;
\item containing links from knowledge base entities to topics, to enable NLP tasks.
\end{itemize}
For these reasons, we see that the choice of topic set can have significant implications for the success of a text labeling, topic modeling, or other NLP applications \cite{Bhargava:lithiumnlp}.

To address these challenges, we present Klout Topics (KT), a curated, lightweight ontology of over 8K nodes and over 13K edges, and discuss:
\begin{itemize}[noitemsep,nolistsep]
\item its characteristics and structure,
\item its relative strengths compared to alternatives,
\item its coverage measured against topics used in Twitter Interests\footnote{\url{http://business.twitter.com/en/targeting/interest.html}}and Google Ads\footnote{\url{https: //support.google.com/ads/answer/2842480?hl=en}}, and 
\item case studies applying it to social media profiles.
\end{itemize}

%% file: texfiles/ontology_problem_statement.tex
\begin{figure}[t]
\centering
\includegraphics[scale=0.15]{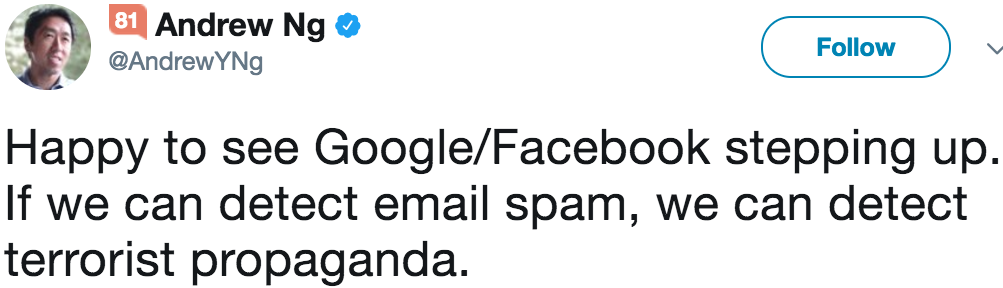}
\caption{Tweet by Andrew Ng, 18 June 2017}
\label{fig:andrewng}
\vspace{-0.1in}
\end{figure}

How can the choice of a candidate topic set impact the success of an interest modeling application? Consider the topics that could be assigned to the Twitter profile of machine learning expert Andrew Ng\footnote{\url{http://twitter.com/AndrewYNg}}. For an interest modeling application to assign meaningful labels to this profile, it would need a topic set containing concepts like ``AI'', ``Coursera'', and ``Stanford''. But is there value in representing ``AI'' and ``Artifical Intelligence'' as distinct topics? Should the topic set contain entities like ``Baidu AI Group'', and would we expect them to be successfully modeled? Should the topic set contain abstract concepts like ``Thought'' or ``Power'', and would those convey any meaning when assigned? If the topic set included negatively-viewed concepts such as ``Terrorist Propaganda'' or ``Spam'', and those topics were assigned based on a post like Figure \ref{fig:andrewng}, would we risk implying that the user is either a terrorist or a spammer? These risks increase when modeling interests over large numbers of social media profiles, and make it advisable to use a carefully curated topic set.

%% file: texfiles/ontology_best_practices.tex
Through the practice of modeling social media interest and expertise topics on Klout.com, we have identified the following best practices for an interest modeling topic set:
\begin{enumerate}[nolistsep,noitemsep]
\item \textbf{Contain concepts that can be meaningful interests of a person or topics of a document.} One can reasonably say ``This social media user has in-depth knowledge of geometry'' or ``this article is about geometry'', which implies that ``Geometry'' is a good candidate topic. However, one cannot reasonably say ``This social media user has in-depth knowledge of rectangles'', which tells us that assigning  ``Rectangles'' as a topic will likely be seen as a failure of the modeling application.
\item \textbf{Be limited to topics, rather than text formats, user demographics, or other attributes}. We are supporting the task of determining what a given piece of text is about and then aggregating that topic knowledge to label a social profile; it is out of our scope to determine non-topical attributes of the user or the text.
\item \textbf{Contain sufficient depth and range of concepts to cover the wide variety of topics discussed on social media.} A greater number of distinct, meaningful topics allows for more meaningful targeting or labeling of users. Narrower subcategories such ``Gospel Music'' or ``Affordable Care Act'' can be more useful to an application than higher-level categories like ``Music'' or ``Politics''. 
\item \textbf{Be small enough for ease of maintenance.} While a too-small topic set has limited utility, as discussed above, a very large topic set is challenging to curate. In addition, a very large topic set can be expensive to model over large amounts of social media content. 
\item \textbf{Be up-to-date with topics likely to be discussed on social media.} Certain domains, such as technology, politics and popular culture, evolve rapidly; a topic set from even a year ago is likely to have problematic gaps.
\item \textbf{Be readable by end users.} If the modeled topics are to be exposed in an application, they should be named with clarity and brevity. (e.g., ``Humidifiers'' is preferable to ``Humidifier (Product)'').
\item \textbf{Be structured hierarchically to leverage connections between concepts}. This allows the modeling application to infer knowledge upward in the graph (e.g., to understand that a user who is interested in ``Chicago Cubs'' will also be interested in ``Baseball'').
\item \textbf{Support multiple languages.} A single-language topic set will limit the final application.
\item \textbf{Include knowledge base entity ids, for data extensibility.} Mapping topics to entities in a knowledge base such as Wikipedia or Freebase is useful for a number of NLP tasks, such as entity disambiguation.
\end{enumerate}

%% file: texfiles/ontology_alternatives.tex
We find that the alternative sources of topic labels fall into the following categories:
\begin{enumerate}[nolistsep,noitemsep]
\item \textbf{Formal ontologies developed for the Semantic Web such as OpenCyc\footnote{\url{http://opencyc.org/}}, OntoLingua \cite{gruber1992ontolingua}, Schema.org\footnote{\url{http://schema.org/}}, etc.} These are larger and more complex than the social media topic domain requires, and would require extra processing to apply. For example, the OpenCyc node ``Game'' is, among other things, an instance of the ``type of temporally stuff-like thing'', which is a complex category out of the scope of topic modeling tasks. 
\item \textbf{Knowledge bases} such as Wikidata\footnote{\url{https://www.wikidata.org/}}, DBPedia\footnote{\url{http://wiki.dbpedia.org/}}, or YAGO\footnote{\url{http://www.mpi-inf.mpg.de/departments/databases-and-information-systems/research/yago-naga/yago/}}. Knowledge bases are by design very large, and contain many entities that are too general, abstract, or specific to be successfully modeled as interests. 
\item \textbf{Lightweight category sets used for ad targeting such as GoogleAds, Twitter Ads, etc.} These are the closest match for the task of interest modeling, and are quite sufficient for many applications; they are the logical basis for comparison for KT. However, because they are designed to be applied in a self-serve UI by advertisers, they tend to be high-level, limited in size, and contain grab-bag or ``Other''-type categories (e.g., ``Fun and Trivia'') that are not present as topics on social media.
\item \textbf{DMOZ\footnote{\url{https://www.dmoz.org/}} or similar URL catalogs.} Because these are intended to catalog URLs and not social media content, they contain format categories that are out of scope of a topic modeling task (e.g. ``Personal Homepages'').
\item \textbf{Yelp or other business catalogs.} These exclude many non-business topics that are present on social media (e.g., ``Donald Trump'', ``History'').
\item \textbf{Amazon, GoodRelations \cite{good-rel} and other product catalogs.} These exclude many non-product topics that are present on social media (e.g., ``Donald Trump'', ``History'').
\item \textbf{LinkedIn or other catalogs of professional skills.} These exclude many non-professional topics that are popular on social media (``Fan Fiction'').
\end{enumerate}

%% file: texfiles/ontology_characteristics.tex
\begin{table}[t]
\centering
\scalebox{0.8}{
\begin{tabular}{|c|c|c|}
 \hline
 \textbf{Root} & \textbf{Nodes} & \textbf{Edges} \\ [0.5ex] 
 \hline
Arts and Humanities & 565 & 625 \\ \hline
Food and Drink & 268 & 313 \\ \hline
Government and Politics & 484 & 527 \\ \hline
Health and Wellness & 374 & 422 \\ \hline
News and Media & 157 & 164 \\ \hline
Science and Nature & 635 & 623 \\ \hline
Sports and Recreation & 827 & 894 \\ \hline
Location & 912 & 1019 \\ \hline
Fashion & 242 & 259 \\ \hline
Education & 648 & 656 \\ \hline
Hobbies & 92 & 110 \\ \hline
Entertainment & 5779 & 5325 \\ \hline
Business & 809 & 884 \\ \hline
Technology & 1131 & 1196 \\ \hline
Lifestyle & 635 & 681 \\ \hline
Travel and Tourism & 73 & 79 \\ [1ex] 
 \hline
\end{tabular}}
\caption{Root categories of KT, with counts of nodes and edges}
\label{table:1}
\vspace{-0.3in}
\end{table}

We present Klout Topics, a lightweight curated ontology 
designed to model user and text topics across social networks. The latest version can be accessed on GitHub\footnote{\url{https://github.com/klout/opendata/tree/master/klout_topic_ontology}}.

\subsection{Methodology}
KT was initially bootstrapped from a set of 140K Freebase topic nodes, selected by matching popular keywords from social media text. That set was curated down to 20K candidates by removing nodes that were not sufficiently popular on social media (``Australian Desert Raisin'') or not sufficiently intuitive for application use (``Comments''). Relationships within those 20K nodes were then inferred from Freebase and verified by curators. With ongoing curation, KT reached its current size of 8K nodes and 13K edges, with newly relevant topics added and obsolete ones removed.

\subsection{Scope}
As users on social networks interact with a variety of topics and domains, KT aims to include any concept that:
\begin{enumerate}[nolistsep,noitemsep]
\item \textbf{is specific enough} to meaningfully describe a user's interest or the primary topic of a piece of text (e.g., ``Life'' or ``Growing'' are too broad, but ``Biology'' or ``Child Development'' should be included).
\item \textbf{is general enough} to be shared by a community of interest on social media, with content that can be detected and recommended. A topic that is theoretically possible, but lacks users or content (e.g. ``Scandinavian Death Metal Played on Kazoos'') will not be included.
\item \textbf{is not redundant} with topics already present (``Child Development Milestones'' does not need to be a separate topic from ``Child Development''). 
\item \textbf{is not illegal or offensive} to a general audience.
\end{enumerate}

\subsection{Components and definitions}

\subsubsection{Topic nodes}

Each topic node in KT contains:
\begin{enumerate}[nolistsep,noitemsep]
\item \textbf{topic\_id}: a unique numerical identifier for the topic. 
\item \textbf{slug}: a human-readable English label, usable in a URL.
\item \textbf{metadata}: includes both the Wikidata entity ID and Freebase machine\_id of the closest corresponding entity. This facilitates NLP tasks such as entity disambiguation and linking  \cite{Bhargava:lithiumnlp}. 
\end{enumerate}

\subsubsection{Topic Edges}
KT is structured as a directed graph 
with all branches connected to one or more of the 16 root categories in Table \ref{table:1}. 
Each edge indicates a parent-child relationship where a child may be a subclass of the parent (e.g., ``Philosophy'' > ``Ethics''), or a notable entity within the topic context (``Philosophy'' > ``Friedrich Nietzsche''). 
A child may have multiple parents in different branches of the graph, and parents may be at different distances from the root.

Each edge contains:
\begin{enumerate}[nolistsep,noitemsep]
\item \textbf{edge\_id}: a unique numerical identifier for the edge.
\item \textbf{source\_topic\_id}: the topic\_id of the parent topic.
\item \textbf{destination\_topic\_id}: the topic\_id of the child topic.
\end{enumerate}


\subsubsection{Topic Display Names}
 
A topic node is associated with the following fields to define how it should be displayed:
\begin{enumerate}[nolistsep,noitemsep]
\item \textbf{topic\_id}: the unique identifier for the topic.
\item \textbf{language}: the language of the given display\_name.
\item \textbf{display\_type}: a flag to indicate whether the topic should be displayed in the given language.
\item \textbf{display\_name}: the most correct and human-readable form of the topic name, in the given language.
\end{enumerate}
In order for KT to be used in consumer-facing applications, its editorial voice strives to be \textbf{transparent} (avoiding neologisms and domain-specific jargon), \textbf{neutral or positive in tone} (avoiding ideological connotations), and \textbf{inclusive} (avoiding derogatory speech and using self-identified terms for groups of people).

\subsection{Internationalization}

A core assumption of KT is that the concept encoded in a node is \textit{not} language- or region-specific, but can be shared across languages. Rather than maintaining separate topic sets for each language, therefore, a single multi-lingual instance can be used. This approach allows for direct comparison of the interests of users in different languages. For flexibility in the application, however, we include a flag to indicate whether a topic should be displayed in a given language.

Currently, the following languages are supported: Arabic (KSA), English (USA), French, German, and Spanish (EU).


%% file: texfiles/ontology_results.tex

\begin{table} [t]
\begin{minipage}[b]{1\linewidth} 
\centering
\resizebox{\textwidth}{!}{%
\begin{tabular}{|c|c|c|c|}
 \hline
 \textbf{Frequent Interest} & \textbf{Rare Interest} & \textbf{Frequent Expertise} & \textbf{Rare Expertise} \\
 \hline
 Social Media & Unisys & Twitter & Unisys \\
  \hline
 Twitter & Andalusia & Music & Andalusia \\
  \hline
 Music & AstraZeneca & Actors & AstraZeneca \\
   \hline
 Actors & Quebec City & Celebrities & Big Band \\
   \hline
 Software & Inbound Marketing & Media & Apulia \\
   \hline
 Television & Category Theory & Sports & Biotelemetry \\
   \hline
 Internet & Dreamworks & YouTube & Infusium \\
   \hline
 Sports & Marketing Automation & Soccer & Inbound Marketing \\
   \hline
 Media & Biotelemetry & Entertainment & Category Theory \\
   \hline
 Celebrities & Objective-J & Singing & Biolage \\
   \hline
 YouTube & Stridex & Internet & Objective-J \\
   \hline
 Entertainment & Cloudera Impala & Television & Stridex \\
   \hline
 Film & Assisted Living & Humor & Dreamworks \\
   \hline
 Humor & Bavaria & Politics & Marketing Automation \\
   \hline
 Politics & Tajikistan & Portugal & Sketch Comedy \\
   \hline
\end{tabular}}
\caption{Frequently and infrequently assigned Klout Topics.}
\label{table:3}
\end{minipage}
\begin{minipage}[b]{1\linewidth} 
\centering
\scalebox{0.75}{
\begin{tabular}{|c|c|c|c|}
 \hline
 & \textbf{KT} & \textbf{Twitter} & \textbf{Google} \\
 \hline
 Top-level & 16 & 25 & 11 \\
 \hline
 Nodes & 8028 & 350 & 2042 \\ 
 \hline
 Depth & 6 & 2 & 9 \\
 \hline
\end{tabular}}
\caption{KT size and depth vs. Twitter and Google interests.}
\label{table:2}
\end{minipage}
\begin{minipage}[b]{1\linewidth} 
\centering
\scalebox{0.77}{
\begin{tabular}{|p{3.5cm}|p{3.5cm}|}
 \hline
 \textbf{Interest} & \textbf{Expertise} \\
 \hline
 Twitter & Twitter \\
   \hline
 YouTube & YouTube \\
   \hline
 Feminism & Singing \\
   \hline
 Singing & Islam \\
   \hline
 Donald Trump & Feminism \\
   \hline
 Islam & Mecca \\
   \hline
 Leadership & Donald Trump \\
   \hline
 MTV & Riyadh \\
   \hline
 Inspiration & MTV \\
   \hline
 Facebook & Justin Bieber \\
   \hline
 One Direction & Inspiration \\
   \hline
 Instagram & Instagram \\
   \hline
 Nature & K-Pop \\
   \hline
 SnapChat & CNN \\
   \hline
 LGBT & LGBT \\
   \hline
\end{tabular}}
\caption{Frequent Klout Topics not in Twitter or Google sets.}
\label{table:4}
\vspace{-0.3in}
\end{minipage}
\end{table}

Here we discuss how the coverage of Klout Topics compares to other topic sets used for modeling user interests. Table \ref{table:2} compares the topical coverage of KT versus Twitter Interests\cite{twitter-ads} and Google ad categories\cite{google-ads}.
While the number of top-level categories is comparable across all three, KT contains the most overall nodes, as well as a significantly deeper structure than Twitter Interests and a comparable one to Google Ads.

In practice, Klout Topics also allows us to model both widely-shared and relatively narrow interests. Table \ref{table:3} shows a sample of Klout Topics assigned to 50M or more profiles out of over 780M scored on Klout.com (``Frequent Interest/Expertise''), as well as a sample of Klout Topics assigned to 10K or fewer (``Rare Interest/Expertise''), as of June 2017. 

Table \ref{table:4} shows popular Klout Topics (assigned to 10M or more profiles as of late June 2017) that are not present in either the Twitter or Google sets. Other examples of topics unique to KT include some brands (``Converse'', ``Pizza Hut''), news-driven concepts (``Affordable Care Act'', ``Criminal Justice Reform''), and narrow subconcepts  (``West Papua'' in addition to ``Indonesia'', ``Aquariums'' in addition to ``Zoos'', ``DevOps'' in addition to ``Programming''). Conversely, Table \ref{table:5} shows a sample of Twitter and Google topics not present in KT. Analysis finds 24 Twitter and 154 Google topics not represented in some form in KT. Many of these missing topics are product subtypes (``Women's Beachwear'') or categories of URLs (``Skins Themes and Wallpapers'').

\begin{table} [t]
\begin{minipage}[b]{1\linewidth} 
\centering
\scalebox{0.78}{
\begin{tabular}{|c|c|}
 \hline
 \textbf{Twitter} & \textbf{Google} \\
  \hline
   Dresses and skirts & ATMs and Branch Locations \\
  \hline
     Empty nesters & Bottled Water \\
  \hline
     Kids' apparel & Clip Art and Animated GIFs \\
  \hline
     Men's beachwear & Men's Interests \\
  \hline
     Venues & Offbeat \\
  \hline
     Women's accessories & Puzzles and Brainteasers \\
  \hline
     Women's beachwear & Song Lyrics and Tabs \\
  \hline
     Women's jeans & Special Occasions \\
  \hline
     Women's pants & Vacation Offers \\
  \hline
     Women's tops & World Localities \\
  \hline
\end{tabular}}
\caption{Sample Twitter and Google interests not in KT.}
\label{table:5}
\vspace{-0.3in}
\end{minipage}
\end{table}



%% file: texfiles/ontology_sample_case_study.tex
\begin{figure}[t]
\centering
\begin{subfigure}{0.45\textwidth}
\includegraphics[scale=0.35]{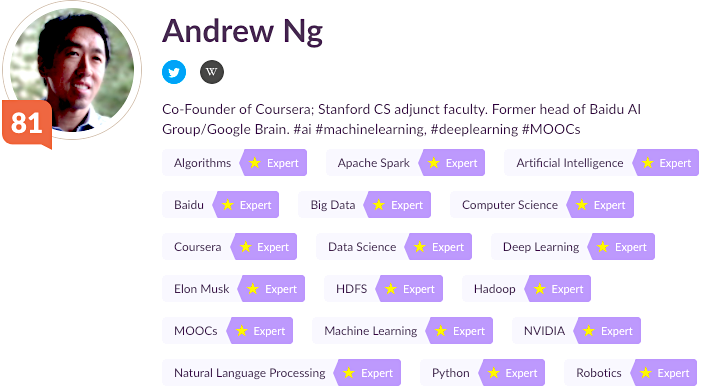}
\caption{Klout Topics on the Klout.com profile of Andrew Ng}
\label{fig:andrewng2}
\end{subfigure} 
\centering
\begin{subfigure}{0.45\textwidth}
\includegraphics[scale=0.35]{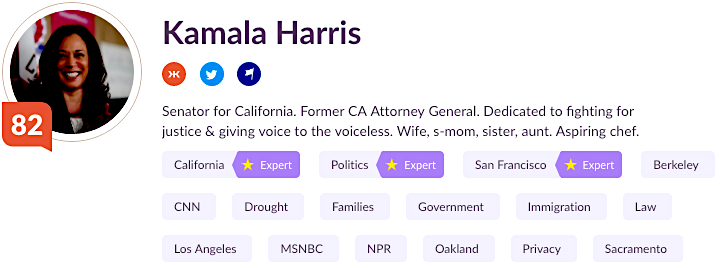}
\caption{Klout Topics on the Klout.com profile of Kamala Harris}
\label{fig:kamalaharris}
\end{subfigure} 
\centering
\begin{subfigure}{0.45\textwidth}
\includegraphics[scale=0.35]{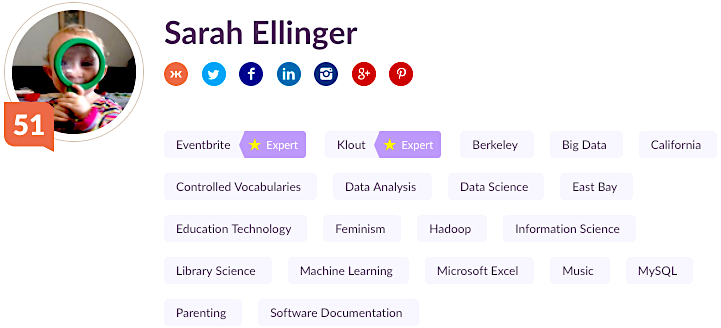}
\caption{Klout Topics on the Klout.com profile of Sarah Ellinger}
\label{fig:sarahellinger}
\vspace{-0.1in}
\end{subfigure} 
\caption{Klout Topics on sample user profiles from Klout.com}
\label{fig:klouitprofiles}
\vspace{-0.2in}
\end{figure}

Here, we show examples of how KT is used to model expertise topics of over 780M social media users on Klout.com, using the Klout topic assignment system (\cite{nemanja-lasta}, \cite{spasojevic2016mining}). These topic scores are also available via the Twitter PowerTrack API\footnote{\url{http://support.gnip.com/apis/powertrack/}}.

Figure [\ref{fig:andrewng2}] shows the expertise topics displayed on the Klout.com profile of Andrew Ng, which range from the broad (``Computer Science'') to the relatively narrow (``Apache Spark''). Figure [\ref{fig:kamalaharris}] shows the Klout.com profile of Senator Kamala Harris, and includes many topics specific to her state of California as well as some relating to government and politics. Figure [\ref{fig:sarahellinger}] shows the Klout.com profile of one of the authors of this paper, and includes a mix of modeled expertise topics (``Eventbrite'', ``Klout'') as well as non-modeled topics from KT that the user has selected (``Parenting'').

%% file: texfiles/ontology_conclusion.tex
In this paper, we have described the importance of a robust topic set for text labeling and topic modeling applications taking input from social media; the topic set should be lightweight, human-readable, span multiple domains of knowledge, and be mapped to a commonly used knowledge base for extensibility to natural language processing tasks. We then discussed the drawbacks of some of the currently available options for this purpose, and proposed Klout Topics (KT) instead. We gave an overview of KT's purpose, characteristics, and coverage; and finally, we gave examples of how it can be used to model a range of topics based on a person's social network activity.